\documentclass[aps,prl,twocolumn,superscriptaddress]{revtex4-1}

\usepackage{hyperref} 
\usepackage{graphicx}
\usepackage{amsmath}
\usepackage{amssymb}
\usepackage{siunitx}

\DeclareGraphicsExtensions{.png,.jpg,.eps}

\usepackage{xcolor}

\begin{document}
\title{Emergence of criticality through a cascade of delocalization transitions in quasiperiodic chains}

\author{V.~Goblot}
\thanks{These two authors contributed equally}
\affiliation{Centre de Nanosciences et de Nanotechnologies (C2N), CNRS, Universit\'{e} Paris-Sud, Universit\'{e} Paris-Saclay, 91120 Palaiseau, France}

\author{A.~\v{S}trkalj}
\thanks{These two authors contributed equally}
\affiliation{Institute for Theoretical Physics, ETH Zurich, 8093 Z\"{u}rich, Switzerland}

\author{N.~Pernet}
\affiliation{Centre de Nanosciences et de Nanotechnologies (C2N), CNRS, Universit\'{e} Paris-Sud, Universit\'{e} Paris-Saclay, 91120 Palaiseau, France}

\author{J.~L.~Lado}
\affiliation{Institute for Theoretical Physics, ETH Zurich, 8093 Z\"{u}rich, Switzerland}
\affiliation{Department of Applied Physics, Aalto University, Espoo, Finland}

\author{C.~Dorow}
\author{A.~Lema\^itre}
\author{L.~Le~Gratiet}
\author{A.~Harouri}
\author{I.~Sagnes}
\author{S.~Ravets}
\affiliation{Centre de Nanosciences et de Nanotechnologies (C2N), CNRS, Universit\'{e} Paris-Sud, Universit\'{e} Paris-Saclay, 91120 Palaiseau, France}

\author{A.~Amo}
\affiliation{Laboratoire de Physique des Lasers Atomes et Mol\'{e}cules (PhLAM), 59000 Lille, France}

\author{J.~Bloch}
\affiliation{Centre de Nanosciences et de Nanotechnologies (C2N), CNRS, Universit\'{e} Paris-Sud, Universit\'{e} Paris-Saclay, 91120 Palaiseau, France}

\author{O.~Zilberberg}
\affiliation{Institute for Theoretical Physics, ETH Zurich, 8093 Z\"{u}rich, Switzerland}

\date{\today}

\maketitle

\textbf{
Conduction through materials crucially depends on how ordered they are. Periodically ordered systems exhibit extended Bloch waves that generate metallic bands, whereas disorder is known to limit conduction and localize the motion of particles in a medium~\cite{anderson1958,akkermans_montambaux_2007}. In this context, quasiperiodic systems, which are neither periodic nor disordered, reveal exotic conduction properties, self-similar wavefunctions, and critical phenomena~\cite{suck2013quasicrystals}. Here, we explore the localization properties of waves in a novel family of quasiperiodic chains obtained when continuously interpolating between two paradigmatic limits~\cite{kraus2012}: the Aubry-Andr\'e model~\cite{aubry1980, jitomirskaya1999}, famous for its metal-to-insulator transition, and the Fibonacci chain~\cite{kohmoto1983, ostlund1983}, known for its critical nature. Using both theoretical analysis and experiments on cavity-polariton devices, we discover that the Aubry-Andr\'e model evolves into criticality through a cascade of band-selective localization/delocalization transitions that iteratively shape the self-similar critical wavefunctions of the Fibonacci chain. 
Our findings offer (i) a unique new insight into understanding the criticality of quasiperiodic chains, (ii) a controllable knob by which to engineer band-selective pass filters, and (iii) a versatile experimental platform with which to further study the interplay of many-body interactions and dissipation in a wide range of quasiperiodic models.
}

Coherent localization of waves is one of the most fundamental effects affecting conduction properties of materials~\cite{akkermans_montambaux_2007}. In pristine periodic mediums, wavelike excitations are expected to propagate ballistically, following their specific wave equation. Commonly, the presence of disorder reduces the wave propagation, possibly driving it to a diffusive instead of a ballistic regime. With increasing disorder in a system, a metal-to-insulator transition occurs and the waves localize. This is known as Anderson's localization transition~\cite{anderson1958, lee1985, akkermans_montambaux_2007}. 
Such disorder effects are found in a variety of physical systems~\cite{RMPAnderson,NatPhysAnderson}. 

Wave propagation in quasiperiodic media is more complex~\cite{suck2013quasicrystals}. These systems are ordered but non periodic, and thus fall in between periodic and randomly disordered systems. The physics of quasiperiodic systems is known to show a plethora of unconventional phenomena such as a one-dimensional (1D) localization transition at a finite (quasi-)disorder strength~\cite{aubry1980, jitomirskaya1999, aulbach_2004}, fractal eigenmodes~\cite{kohmoto1983, ostlund1983}, and critical behavior~\cite{jitomirskaya1999, suck2013quasicrystals}. Among the variety of quasiperiodic models, two canonical examples are the Aubry-Andr\'{e} (AA)~\cite{aubry1980, jitomirskaya1999} and the Fibonacci model~\cite{kohmoto1983, ostlund1983}, which are currently drawing much attention, in particular, in the context of many-body localization~\cite{mastropietro2015,schreiber2015,bordia2017,mace2019,varma2019}.
The quasiperiodicity of the AA model enters in the form of an on-site cosine modulation incommensurate with the underlying periodic lattice spacing, whereas the Fibonacci model has a modulation with two discrete values that appear interchangeably according to the Fibonacci sequence. 
Interestingly, the AA and the Fibonacci modulations have very different localization properties.
Specifically, the AA model shows a unique self-dual localization transition~\cite{aubry1980, jitomirskaya1999}, whereas the Fibonacci model always has critical wave functions~\cite{kohmoto1983,ostlund1983}.
Recently, it has been shown that these two paradigmatic models belong to the same topological class and can be viewed as two limits of an interpolating Aubry-Andr\'{e}-Fibonacci (IAAF) model~\cite{kraus2012, Verbin2013, Verbin2015}. The IAAF model has been useful for the description of the topological properties of Fibonacci chains~\cite{kraus2012,Verbin2013,kraus2016quasiperiodicity}, and for generating topological pumps~\cite{kraus2012a, Verbin2015}. The IAAF model also provides a unique playground to explore how criticality develops during a smooth interpolation between the AA and Fibonacci models.

\begin{figure*}[t!]
	\centering
    \includegraphics[width=\textwidth]{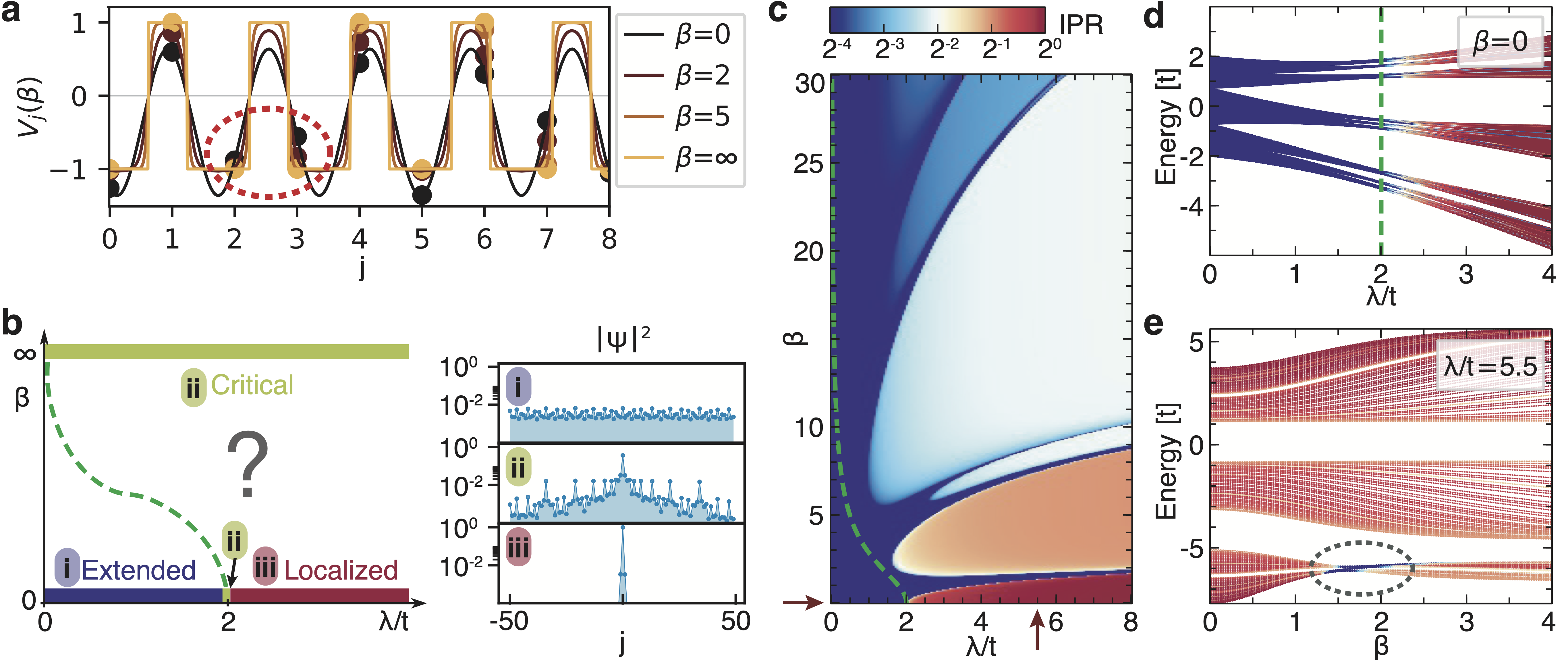}
    \caption { \textbf{Interpolating Aubry-Andr\'e-Fibonacci modulation and theoretical localization phase diagram.}
	\textbf{a,} 
	    Spatial on-site potential [Eq.~\eqref{pot}] evaluated for several values of $\beta$. The two limiting cases are (black line) the Aubry-Andr\'{e} (AA) modulation for $\beta=0$  and (orange line) the Fibonacci modulation for $\beta=\infty$. Pair of states circled in red are nearest neighbors in space and are close in energy such that they hybridize with a finite hopping strength.   
	\textbf{b,} 
	    (left) Localization phase diagram explored in this paper. At $\beta = 0 $, the AA localization transition occurs at $\lambda / t = 2$, while for $\beta=\infty$ eigenmodes are always critical.
	    Dashed (green) line, also shown in \textbf{c}, marks the decrease in the extended region obtained using a generalized self-duality argument (see Supplementary Section 1). 
	    (right) Typical spatial distribution for (i)  extended (ii)  critical and (iii) localized modes. 
	\textbf{c,} 
	    IPR [cf.~Eq.~\eqref{ipr}] of the lowest eigenmode of the tight-binding model [Eq.~\eqref{ham}] as a function of $\beta$ and  $\lambda$. Red arrows mark $\beta$ and $\lambda$ values corresponding to \textbf{d} and \textbf{e}, respectively.
    \textbf{d,}	
        The IPR of all eigenmodes of  Eq.~\eqref{ham} as a function of energy and $\lambda/t$ for $\beta=0$, i.e., in the AA limit. The dashed (green) line marks the critical point at $\lambda/t=2$.	   
    \textbf{e,} 
        The IPR of all eigenmodes of  Eq.~\eqref{ham} as a function of energy and $\beta$ for $\lambda/t=5.5$. At $\beta \sim 1.5$, the lowest energy set of eigenmodes become extended (dashed circle). \textbf{c-e}, We evaluate the IPR on a chain with $L=610$ sites.
        }
        \label{fig1}
\end{figure*}

In this work, we investigate, both theoretically and experimentally, the localization phase diagram of the IAAF model. We show that, along the continuous deformation of the AA into a Fibonacci model, eigenmodes undergo a cascade of band-selective localization/delocalization transitions. We report an experimental observation of such transitions using polaritonic one-dimensional (1D) chains, where we take advantage of the fact that sculpting polaritonic wires is particularly suitable for direct imaging of the modes both in real and reciprocal space in complex potential landscapes.
With our combined theoretical and experimental analysis, we identify the mechanism behind the cascade of localization/delocalization transitions: the metallic region of the AA model gradually shrinks as the potential becomes steeper (more pronounced) when the AA morphs into the Fibonacci; the cascade of transitions involves hybridization of localized modes that thus gradually extend to become critical in the limit of the Fibonacci model. Interestingly, the cascade to criticallity appears in quantized plateaus that gradually increase the eigenmode localization length. Moreover, the band-selective delocalization offers a mechanism by which to engineer band-pass filters, and puts forward a promising platform to explore the interplay between quasiperiodicity and many-body interactions.

The IAAF model~\cite{kraus2012, Verbin2013, Verbin2015} is a 1D tight-binding chain with a quasiperiodic on-site potential modulation
	\begin{equation} \label{ham}
		\mathcal{H} \, \psi_j = t \, ( \psi_{j+1} + \psi_{j-1} ) + \lambda \,  V_j(\beta) \, \psi_{j}\,,
	\end{equation}
where $\psi_j$ is the wave function at site $j$, $t$ the nearest-neighbour hopping amplitude, and $\lambda$ the amplitude of the on-site potential modulation. The on-site potential (see Fig.~\ref{fig1}a) is defined as
	\begin{equation} \label{pot}
		V_j (\beta) = -\frac {\tanh {\beta [\cos{(2\pi b j + \phi)} -\cos{(\pi b)} }]} {\tanh \beta}\,,
	\end{equation}
with the spatial modulation frequency taken as the inverse of the golden mean, $b = 2/(\sqrt{5}+1)$. Since the frequency $b$ is irrational, the potential is incommensurate with the underlying lattice and the model is quasiperiodic. The parameter $\phi$ acts as a global spatial shift of the potential and although crucial for many effects, such as topological pumping~\cite{kraus2012a, Verbin2015,kraus2016quasiperiodicity}, it does not affect the localization properties. The tunable parameter $\beta$ provides a knob by which to interpolate between two known limiting cases: (i) $\beta \rightarrow 0$ reduces to the AA modulation~\cite{aubry1980, harper1955}, up to a constant energy shift $V_j^{\rm AA} (\beta) = \cos{(2\pi b j + \phi)} -\cos{(\pi b)}$, and (ii) $\beta \rightarrow \infty$ corresponds to a step potential switching between $\pm 1$ values according to the Fibonacci sequence~\cite{kohmoto1983, ostlund1983}.

\begin{figure*}[t!]
    \centering
    \includegraphics[width=\textwidth]{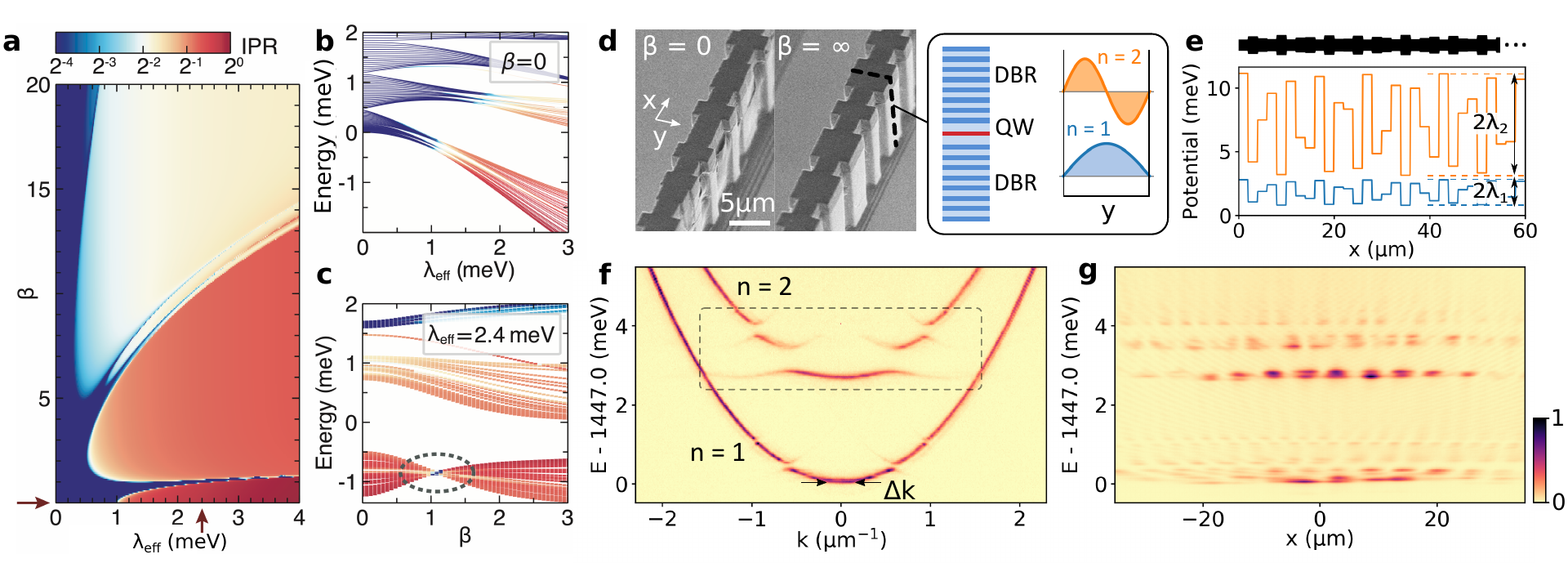}
    \caption{ \textbf{Continuum IAAF model and its experimental implementation.}
    \textbf{a,} 
	    The IPR [cf.~continuum version of Eq.~\eqref{ipr} in Supplementary Section~4] for the lowest-energy eigenmode of the continuum IAAF model [Eq.~\eqref{nf_hamiltonian}].
	    Two red arrows mark the constant $\beta$ and $\lambda_{\rm eff}$ values used in \textbf{b} and \textbf{c}, respectively. 
    \textbf{b,} 
        The IPR of all eigenmodes of Eq.~\eqref{nf_hamiltonian} as a function of energy and $\lambda_{\rm eff}$ for $\beta=0$ (AA limit). 
    \textbf{c,} 
        The IPR of all eigenmodes of Eq.~\eqref{nf_hamiltonian} as a function of energy and $\beta$ for $\lambda_{\rm eff}=2.4$ meV. The region where the lowest band is delocalized is marked with a dashed circle.
\textbf{d,}
        Scanning electron micrograph of two modulated polariton wires implementing the IAAF model for $\beta=0$ and $\beta = \infty$. (inset) Schematic representation of the cavity structure along the $z$-direction, with a single quantum well (QW) embedded between two distributed Bragg reflectors (DBR). On the right, the transverse ($y$-direction) profile of the $n=1$ and $n=2$ polariton subband are depicted; 
	\textbf{e,} 
	    The IAAF potential for the $n=1$ [$n=2$] polariton subband (blue [orange] lines) corresponding to a modulation of the same wire section shown on top (black contour). 
    \textbf{f,}
    	Photoluminescence (PL) intensity measured as a function of momentum $k$ and energy for a wire corresponding to  $\beta=0$ and $\lambda_1 = 0.2$ meV. The $n = 1$ and $n = 2$ sets of subbands are identified.
	\textbf{g,} 
	    PL intensity measured as a function of space $x$ and energy for the same wire as in \textbf{f}.
	 }
	\label{fig2}
\end{figure*}

Unlike standard Anderson localization under on-site disorder~\cite{anderson1958}, the localization transition for the AA model ($\beta=0$) occurs for all eigenmodes at the same nonzero critical point~\cite{aubry1980, jitomirskaya1999} (see $x$-axis of Fig.~\ref{fig1}b). The critical point can be obtained using a self-duality argument~\cite{aubry1980, jitomirskaya1999}: for $\lambda/t<2$, all modes are extended, for $\lambda/t>2$ they are localized, and at the critical point $\lambda/t=2$, all the modes are critical and self-similar with a power-law spatial decay, see Fig.~\ref{fig1}b for representative modes in the three scenarios. In the limit of $\beta\rightarrow\infty$, all the eigenmodes of the Fibonacci model are critical for any finite $\lambda/t>0$~\cite{kohmoto1983, ostlund1983}. 
The main goal of this work is to explore the IAAF localization phase diagram and understand how AA modes continuously develop into critical Fibonacci modes. Note that previous studies of deformations of a cosine potential into a step function observed the appearance of band edges but did not reach the critical Fibonacci model~\cite{hiramoto1989}. Crucially, the IAAF \eqref{pot} contains a constant energy shift, $\cos(\pi b)$, that guarantees the correct Fibonacci limit ($\beta\rightarrow\infty$).

We first develop an intuitive picture of what we expect to observe: as $\beta$ increases, the potential becomes steeper (see Fig.~\ref{fig1}a), and effectively should lead to stronger localization, i.e., the region where the modes are extended shrinks (see Fig.~\ref{fig1}b). More precisely, we theoretically investigate the transition to criticality by computing the eigenmodes of  Eq.~\eqref{ham} and systematically analysing the inverse participation ratio (IPR) of each eigenstate $\psi_n$ as a measure of its localization,
	\begin{equation} \label{ipr}
		{\rm IPR}_n \equiv \frac{\sum_{j=1}^{L} |\psi_{n,j}|^4}{\sum_{j=1}^{L}|\psi_{n,j}|^2} \, ,
	\end{equation}
where the sums run over length $L$ of the chain. In the regime where the $n^{\rm th}$ eigenmode $\psi_n$ is extended, the IPR is equal to the inverse of the system length (${\rm IPR}_n = 1/L$) and drops to 0 for an infinite system. Conversely, for modes localized on $N$ sites, the IPR is equal to $1/N$ and remains finite for infinite system size. 

In Figs.~\ref{fig1}c-e, we summarize the IPR values obtained within the tight-binding analysis. Let us start with Fig.~\ref{fig1}d, which illustrates the spectral dependence of the IPR for $\beta=0$. The AA localization transition, occurring simultaneously for all eigenmodes at $\lambda/t=2$, is clearly seen.  Fig.~\ref{fig1}c shows the IPR of the lowest energy eigenmode as a function of the IAAF parameters $\lambda/t$ and $\beta$. 
The IPR does not evolve monotonously with $\beta$ but presents a cascade of lobes of higher IPR values (red regions in Fig.~\ref{fig1}c) separated by minima of IPR (blue regions in Fig.~\ref{fig1}c). Similar lobe structures occur for all eigenmodes (see Supplementary Section~2). At low $\lambda/t$, when increasing $\beta$, the region where the eigenmode is extended decreases, as expected from the steeper potential (see Supplementary Section~1). We now focus on the cascade to criticality for $\lambda/t=5.5$, i.e., starting from the strongly localized AA and continuously evolving toward the Fibonacci limit. As summarized in Fig.~\ref{fig1}e, we observe that the lowest set of eigenmodes squeeze into a narrow spectral window, hybridize due to the finite hopping strength $t$, and delocalize at $\beta \sim 1.5$. By further increasing $\beta$, the modes localize once more with a smaller IPR. 
This process repeats at each minimum of the IPR, see Fig.~\ref{fig1}c. Furthermore, different bands exhibit this cascade at different values of $\beta$ (see Supplementary Section~2). We conclude that the transition to criticality does not happen uniformly, but instead occurs through successive localization-delocalization transitions. 

\begin{figure*}[t!]
	\centering
	\includegraphics[width=\textwidth]{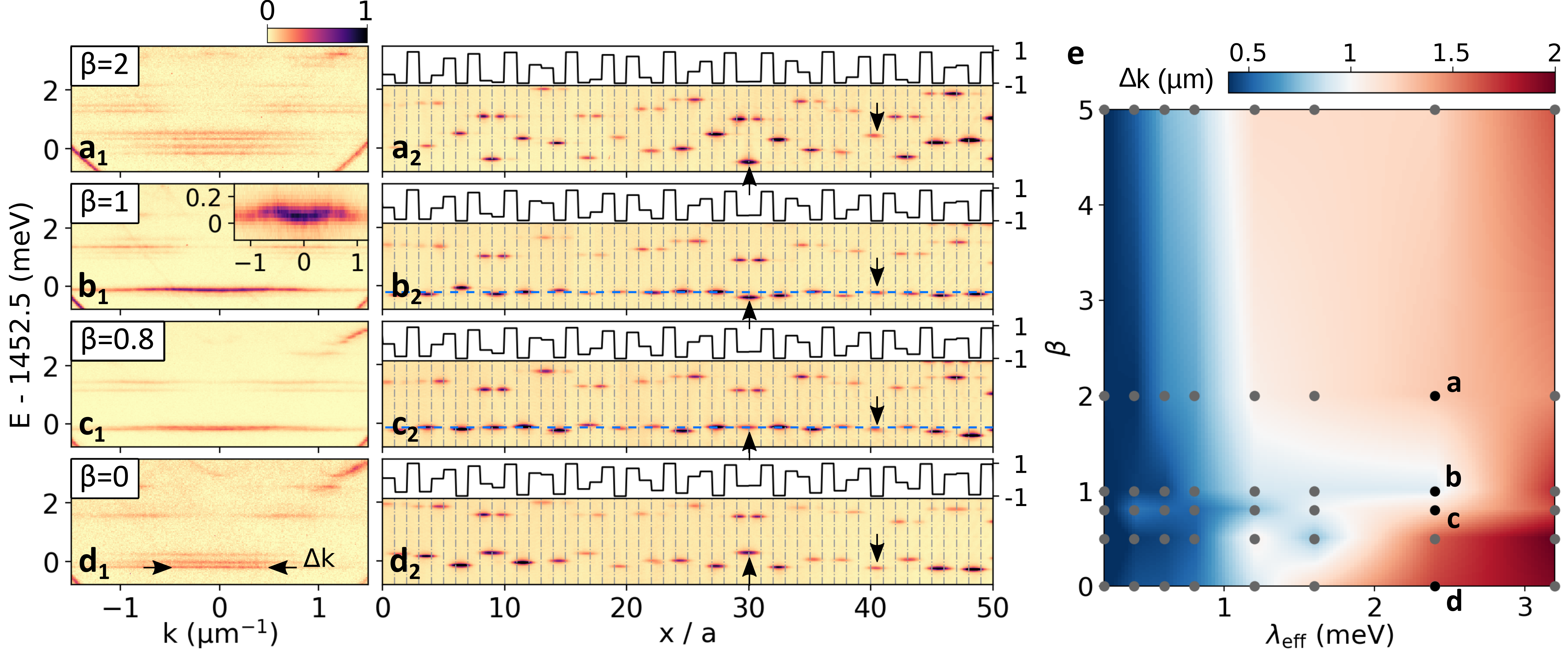}
	\caption{ \textbf{Experimental localization phase diagram.}
		$\boldsymbol{a_1}$-$\boldsymbol{d_1,}$ 
			Photoluminescence (PL) intensity as a function of energy and $k$, zoomed on the $n=2$ modes (cf.~Figs.~\ref{fig2}f and g), for $\lambda_2 = \SI{2.4}{\meV}$ and $\boldsymbol{a_1}$: $\beta=2$, $\boldsymbol{b_1}$: $\beta=1$, $\boldsymbol{c_1}$: $\beta=0.8$ and $\boldsymbol{d_1}$: $\beta=0$, as indicated in \textbf{e}. Inset in $\boldsymbol{b_1}$ shows a zoom on the lowest energy band to highlight its curvature. 
		$\boldsymbol{a_2}$-$\boldsymbol{d_2,}$ 
			Corresponding real-space energy-resolved PL intensity. Vertical gray-dashed lines indicate the letter edges. On top of each panel, the nominal potential along the wire is depicted. Horizontal (light blue) line marks the energy where the band hybridization takes place. Up (down) arrows mark the modes that invert their energies with $\beta$, which extend over two (one) letters. Modes localized on two letters have clear anti-bonding modes above them.
		\textbf{e,} 
			(grey dots) Full width at half maximum (FWHM), $\Delta k$, of the lowest energy state (as marked with arrows on panel $\boldsymbol{d_1}$) measured for several values of $\lambda_{\rm eff}$ and $\beta$. Colours are obtained by interpolation between measured points.
	}
	\label{fig3}
\end{figure*}

To observe experimentally the aforementioned localization-delocalization transitions, we engineer cavity polariton samples. This photonic platform has been recently used for the exploration of Fibonacci chains: log-periodic oscillations of the density of states and direct measure of topological invariants could be revealed by optical spectroscopy~\cite{Tanese2014, Baboux2017}. The quasiperiodic potential can be treated as a perturbation to the one-dimensional motion of free polaritons, namely, the Hamiltonian of this system  can be written as a continuum model
\begin{equation} \label{nf_hamiltonian}
		\mathcal{H} \, \psi (x) =\left[ - \frac{\hbar^2}{2m} \nabla^2 + \lambda_{\rm eff} V(x, \beta) \right] \psi(x)
\end{equation}
where $m$ is the polariton mass. The modulation $V_j (\beta)$, given by Eq.~\eqref{pot}, is implemented with a piecewise 1D potential $V(x, \beta) = V_{\lfloor x/a \rfloor}(\beta)$, with steps of length $a$. In Figs.~\ref{fig2}a-c, we report the calculated IPR values obtained within the continuum IAAF model~\eqref{nf_hamiltonian} (see Supplementary Section~4). For $\beta=0$ (Fig.~\ref{fig2}b), we observe signatures of the AA localization as a function of $\lambda_{\rm eff}$. Note that contrary to the tight-binding model, the localization does not occur simultaneously for all modes and mobility edges appear in the spectrum~\cite{biddle2011,ganeshan2015, luschen2018, Roati2008}. For the lowest band, the localization transition occurs at $\lambda_{\rm eff} \approx \SI{1}{meV}$, approximately at twice the relevant kinetic energy scale in the band (see Supplementary Section~5).
Importantly, as reported in Figs.~\ref{fig2}a and c, the continuum model also exibits the lobes of localization-delocalization transitions.
Thus either of the two frameworks can be used for experiments.

We fabricated laterally modulated photonic wires based on polariton semiconductor microcavities. The cavity sample is grown by molecular beam expitaxy and consists of a quantum well that is inserted between two high-reflectivity Bragg mirrors along the $z$-direction (see Methods for further details). We process the cavity sample into quasi-1D microstructures using electron beam lithography and dry etching. The photonic modes (also called polaritons) form 1D-subbands with distinct transverse spatial distribution, see Fig.~\ref{fig2}d. The lowest energy subband ($n=1$) presents modes with a maximum at the middle of the wire, while the $n=2$ subband shows modes with intensity maxima left and right of the wire center (see Supplementary Section~6).
For a given transverse mode $n$, the lateral confinement energy for polaritons is given by
$U(w) = (\hbar^2 \pi^2)/(2m) \times n^2/w^2$~\cite{Tanese2014},
with $w$ the width of the wire.
To implement the piecewise potential of the IAAF model \eqref{nf_hamiltonian}, we consider etched sections (dubbed letters) of fixed length $a = \SI{2}{\micro \metre}$ and design their width $w_j$ so that $U(w_j) = U_0 + \lambda_n V_j$, with $U_0$ a global offset determined by $U({\rm max}(w_j))$, see Fig.~\ref{fig2}e. 
Interestingly, due to the proportionality of $U(w)$ with $n^2$, the modulation amplitude $\lambda_n$ for higher-energy subbands is increased by a factor $n^2$ with respect to the $n=1$ subband. It is thus possible to access larger values of $\lambda_{\rm eff} = n^2 \lambda_1$ when considering higher-energy subbands.

To explore the localization properties of polariton modes in these IAAF chains, we optically excite single wires cooled down to 4K using a weak non-resonant continuous-wave laser. The excitation spot is elongated along the wires and we analyze the spectrally resolved photoluminescence (PL) signal either in real or in momentum space (see Methods for further details). In Figs.~\ref{fig2}f and g, we show an example of such measurements for $\beta=0$ and $\lambda_1 = \, 0.2$ meV. Polariton subbands corresponding to $n=1$ and $n=2$ are clearly resolved. The lateral modulation results in the opening of minigaps, which are 4 times larger for the $n=2$ than for the $n=1$ subbands, as expected. In the depicted example, all the polariton modes are extended in real space (see Fig.~\ref{fig2}g), indicating that this particular wire implements a metallic phase for the AA model. 

Let us now discuss polariton localization properties when increasing the value of $\beta$. In Figs.~\ref{fig3}a-d, we present PL measurements in real and reciprocal space for 4 values of $\beta$ and constant $\lambda_{\rm eff}$. For clarity, these figures only show the $n=2$ sub-band that experiences $\lambda_2 = 2.4$~meV. For $\beta=0$ (Fig.~\ref{fig3}d), we observe localized emission spots in real space that are dispersed in energy. Accordingly, the emission is very broad in $k$-space. These features are characteristic of the AA localized phase.
When $\beta$ increases, we observe the merging of lowest-energy emission spots within a narrow spectral window. For $\beta=1$, the k-space image reveals the formation of a band with a finite curvature (see Supplementary Section~7). This clearly indicates the formation of extended modes. By further increasing $\beta$, the polariton modes are localized once more. Thus, these measurements provide evidence for the first delocalization/localization transition when deforming the AA model into the Fibonacci chain. Every detail of the measured spatial patterns shown in Figs.~\ref{fig3}a2-d2 is reproduced by the continuum model simulations (see Supplementary Section~8).
Similar PL measurements were performed on different wires implementing various values of $\beta$ and $\lambda_{\rm eff}$. 
To quantify the polariton localization and obtain the phase diagram, we extract, from the measurements in $k$-space, the full width at half maximum $\Delta k$ of the lowest energy modes in the considered subband. Extended modes have lower $\Delta k$ than localized ones. In Fig.~\ref{fig3}e, we summarize all the measured $\Delta k$ values, and clearly trace the first delocalization lobe in the phase diagram, in agreement with theoretical predictions (cf.~Fig.~\ref{fig1}c and Fig.~\ref{fig2}a). 

\begin{figure}
	 \centering
	 \includegraphics[scale=1]{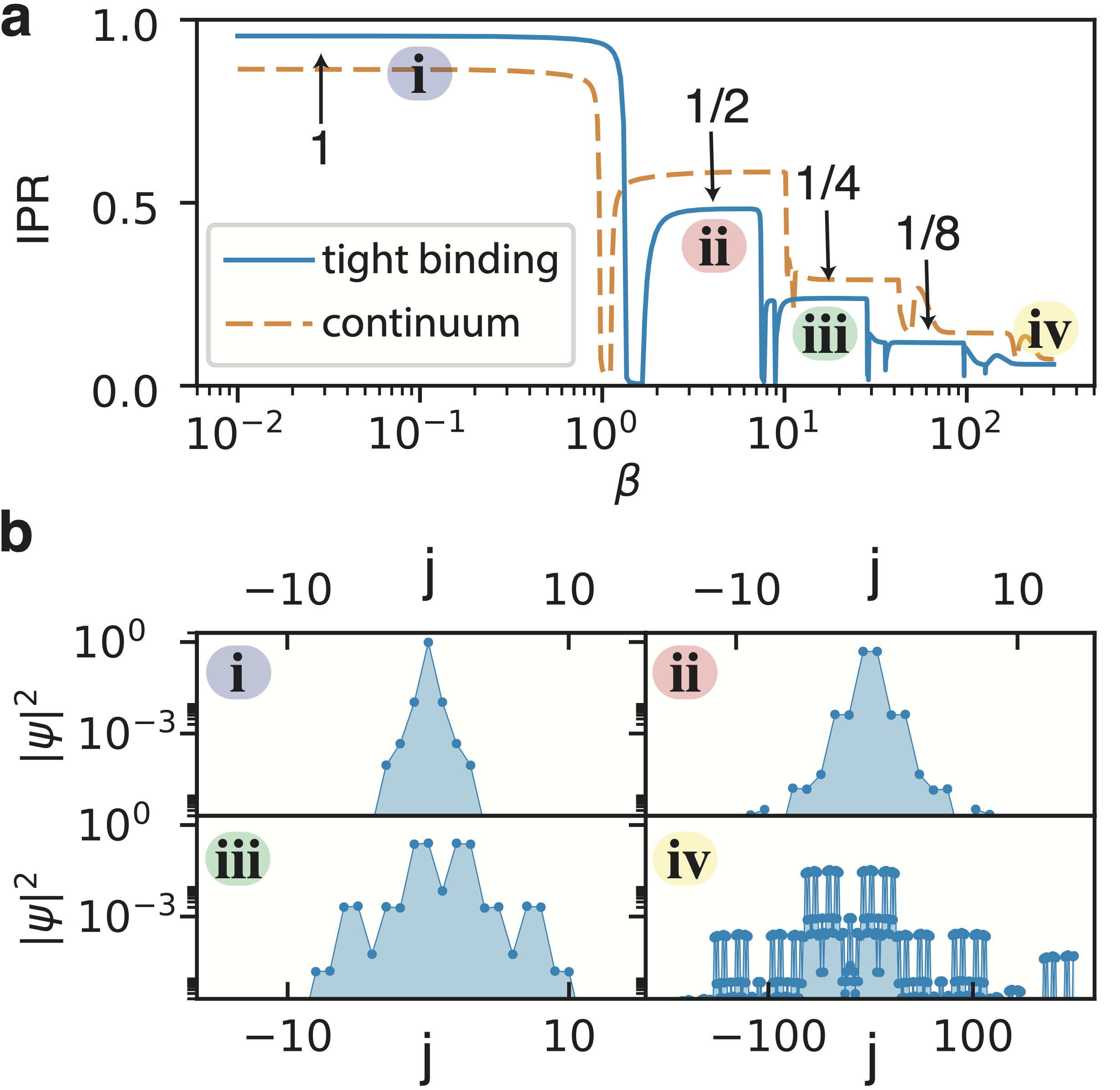}
	 \caption{ \textbf{Spatial evolution with $\beta$ of the lowest energy eigenstate.}
	 \textbf{a},
	    The IPR [cf.~Eq.~\eqref{ipr} and its continuum version in Supplementary Section~4] of the lowest-energy eigenmode as a function of $\beta$ calculated (solid blue line) for $\lambda/t = 5.5$  with the tight-binding model [Eq.~\eqref{ham}], and (dashed orange line) for $\lambda_{\rm eff} = 2.4$ meV in the continuum model [Eq.~\eqref{nf_hamiltonian}].
	    The plateaus correspond to the localized regions of the phase diagram on Fig.~\ref{fig1}\textbf{c}. 
	 \textbf{b},
	    Spatial profile of the lowest-energy eigenmode of tight-binding model for $\lambda/t=5.5$ and (i-iv) $\beta=\{ 0.01,5,15,1000 \}$, respectively.
	\label{fig4}
	}
\end{figure}

Finally, we provide a physical understanding of the cascade of localization transitions discovered in the IAAF model. Careful analysis of the real-space PL images (Figs.~\ref{fig3}a$_2$-d$_2$) reveals a key mechanism in the transition to criticality. In the AA limit (Fig.~\ref{fig3}d$_2$), we can see that all the lowest energy modes are localized within one letter (see, e.g., arrow pointing downwards). Additionally some modes, at higher energies, are localized on two sites corresponding to a two-letter potential minimum (see, e.g., arrow pointing upwards). These modes can be viewed as a bonding hybridization of two single-letter modes and are easily identified by the presence of a high energy anti-bonding mode at the same spatial location. When increasing $\beta$, these two-letter bonding modes decrease in energy. They become resonant with the single-letter modes at the delocalization transition and then become the new lowest energy state for larger $\beta$ (see arrows in Fig.~\ref{fig3}a2-d2). 

This spatial feature can be fully understood within the tight-binding approach. Here, starting at $\beta=0$ (AA limit) and high $\lambda$, the spatial localization length of eigenmodes is known analytically to be $\xi = (\log\lambda/2t)^{-1}$~\cite{aubry1980, jitomirskaya1999}. Thus, for $t\rightarrow 0$, all eigenmodes are expected to be localized on a single site. Nevertheless, the golden mean quasiperiodic modulation guarantees that there always exist pairs of modes that are spatially nearest neighbours and close in energy (see modes marked with red dashed circle in Fig.~\ref{fig1}a and nearly identical neighbouring two-letter potentials in Figs.~\ref{fig3}a2-d2). These neighbouring modes hybridize with any finite hopping strength $t$, and are tuned with $\beta$ to eventually overtake the role of the lowest energy eigenmodes (see Fig.~\ref{fig4}, Supplementary Section~3 and \cite{supmat_video_1}). As a result the IPR is expected to be reduced by a factor of two across the transition.

Crucially, the mechanism of delocalization followed by relocalization repeats itself by further increasing $\beta\rightarrow\infty$. As seen in Fig.~\ref{fig4}a, the IPR for both tight-binding~\eqref{ham} and continuum~\eqref{nf_hamiltonian} models display a series of plateaus whose height decreases in a stepwise fashion. The former model exhibits steps that decrease with a factor of two every time a delocalization transition occurs. The latter shows deviations from this quantization due to the spatial extent of the continuum wavefunctions within a single letter (cf.~continuum model IPR in Supplementary Section~4).
The spatial distribution of the localized eigenmodes on different plateaus is reported in Fig.~\ref{fig4}b. We thus reveal that the transition to criticality at $\beta=\infty$, where the eigenmodes are self-similar, develops in the IAAF through a unique iterative process of eigenmodes hybridization that doubles the spatial extent of the eigenmodes (see also \cite{supmat_video_2}). Further study of the model is aimed at obtaining an analytical expression for this unique transition into criticality.

To conclude, our work reports a new mechanism of the localization of waves in paradigmatic quasiperiodic models. 
The interpolation between the Aubry-Andr\'{e} and Fibonacci models yields a cascade of localization/delocalization transitions, where at each transition the spatial extent of modes doubles followed by a plateau in their localization phase diagram. The transitions do not occur simultaneously for all eigenmodes, but at different values of $\beta$ for different energy bands. These controlled band-selective localization/delocalization transition provide a tunable knob by which to engineer selective band-pass filters. This approach opens up a new frontier where generalizations of the mechanism to other models, with different modulation frequencies and interpolations can be addressed. The high-precision of the polaritonic platform offers effective means to realize these complex potential profiles, and explore their wave localization properties. Furthermore, it uniquely promotes the study of quasiperiodic physics under the influence of non-Hermiticy and of non-linearities on wave localization.

\section*{Acknowledgments}
We thank Y.~E.~Kraus and Y.~Lahini for fruitful discussions. 
 A.~\v{S}. and O.~Z. acknowledge financial support from the Swiss National Science Foundation through grant PP00P2 163818. J.~L.~L. acknowledges financial support from the ETH Fellowship program. This work was supported by the ERC grant Honeypol, the H2020-FETFLAG project PhoQus (820392), the QUANTERA project Interpol (ANR-QUAN-0003-05), the French National Research Agency project Quantum Fluids of Light (ANR-16-CE30-0021), the Paris Ile-de-France R\'egion in the framework of DIM SIRTEQ, the French government through the Programme Investissement d’Avenir (I-SITE ULNE / ANR-16-IDEX-0004 ULNE) managed by the Agence Nationale de la Recherche, the French RENATECH network, the Labex CEMPI (ANR-11-LABX-0007), the CPER Photonics for Society P4S and the M\'etropole Europ\'eenne de Lille (MEL) via the project TFlight.
\section*{Author contributions:}
A.~\v{S}. and J.~L.~L. performed the tight-binding theoretical work; V.~G. and
N.~P. developed the continuum-model simulations; V.~G. and C.~D.
designed the samples; A.~L., L.~L.~G., A.~H. and I.~S. fabricated the
samples; V.~G., N.~P. and C.~D. performed the experiments; V.~G., A.~\v{S}.,
 N.~P., J.~L.~L., S.~R., A.~A., J.~B. and O.~Z. contributed to the data
analysis (simulations and experiments), scientific discussions and to
the writing of the manuscript; J.~B. and O.~Z. supervised the work. \\

{\bf Competing interests:} The authors declare no competing interests. 
{\bf Data and materials availability:} The data presented in this work are available from the corresponding authors upon request. \\
{\bf Correspondence and requests for materials} should be addressed to J.B. (jacqueline.bloch@c2n.upsaclay.fr) and O.Z. (odedz@phys.ethz.ch)

\bigskip
	 
\section{Methods}

\subsection{Sample description} 

The quasiperiodic structures used in this work are etched out of a planar semiconductor microcavity with high quality factor ($Q \approx 75,000$) grown by molecular beam epitaxy. The microcavity consists of a $\lambda$ GaAs layer embedded between two $\mathrm{Ga_{0.9}Al_{0.1}As/Ga_{0.05}Al_{0.95}As}$ distributed Bragg reflectors (DBR) with 36 (top) and 40 (bottom) pairs. A single 8~nm $\mathrm{In_{0.05}Ga_{0.95}As}$ QW is inserted at the center of the cavity, resulting in the strong exciton-photon coupling, with an associated 3.5 meV Rabi splitting.
After the epitaxy, the sample is processed with electron beam lithography and dry etching to form one-dimensional wires with modulated width. The modulation consists of sections of fixed length $a = \SI{2}{\micro \metre}$ and different width $w_j$, designed to implement the IAAF potential $U(w_j)$, with chosen $(\lambda_1, \beta)$. The maximum section width is fixed to $\SI{4}{\micro \metre}$, corresponding to the minimum of the effective 1D potential.

The exciton-photon detuning, defined as energy difference between the uncoupled planar cavity mode and the exciton resonance, is of the order of $\delta = \SI{-20}{\milli \eV}$ for all the experiments. 

\subsection{Experimental technique}

Non-resonant photoluminescence measurements were realized with a single-mode continuous-wave (cw) laser at 780 nm. 
The elongated spot was engineered using a cylindrical lens. The emission was collected through a microscope objective with NA 0.5 and imaged on the entrance slit
of a spectrometer coupled to a charge-coupled device (CCD) camera with a spectral resolution of $\sim \SI{30}{\micro \eV}$. Real- and momentum-space photoluminescence
images were realized by imaging the sample surface and the Fourier plane of the objective, respectively. A polarizer was used to select
emission polarized either along or across the long axis of the chains. The sample was cooled to $T = \SI{4}{K}$.


%


\newpage
\cleardoublepage
\setcounter{figure}{0}
\renewcommand{\figurename}{Supplementary Figure}

\onecolumngrid
\begin{center}
	\textbf{\normalsize Supplemental Material for}\\
	\vspace{3mm}
	\textbf{\large Emergence of criticality through a cascade of delocalization transitions in a quasi-crystal}\\
	\vspace{4mm}
	{ V.~Goblot${}^{1,*}$, A.~\v{S}trkalj${}^{2,*}$, N.~Pernet${}^1$, J.~L.~Lado${}^{2,3}$, C.~Dorow${}^1$, A.~Lema\^itre${}^1$, L.~Le~Gratiet${}^1$, A.~Harouri${}^1$, I.~Sagnes${}^1$, S.~Ravets${}^1$, A.~Amo${}^3$, J.~Bloch${}^1$ and O.~Zilberberg${}^2$ }\\
	\vspace{1mm}
	\textit{\small ${}^1$ Centre de Nanosciences et de Nanotechnologies (C2N), CNRS, Universit\'{e} Paris-Sud, Universit\'{e} Paris-Saclay, 91120 Palaiseau, France}\\
	\textit{\small ${}^2$ Institute for Theoretical Physics, ETH Z\"urich, 8093 Z\"urich, Switzerland}\\
	\textit{\small ${}^3$ Department of Applied Physics, Aalto University, Espoo, Finland}\\
	\textit{\small ${}^4$ Laboratoire de Physique des Lasers Atomes et Mol\'{e}cules (PhLAM), 59000 Lille, France}
	
	\vspace{5mm}
	*These authors contributed equally to this work.
\end{center}
\setcounter{equation}{0}
\setcounter{section}{0}
\setcounter{figure}{0}
\setcounter{table}{0}
\setcounter{page}{1}
\makeatletter
\renewcommand{\bibnumfmt}[1]{[#1]}
\renewcommand{\citenumfont}[1]{#1}

\setcounter{enumi}{0}
\renewcommand{\theequation}{\Roman{enumi}.\arabic{equation}}

\bigskip
\section{I. Self-duality argument}
\setcounter{enumi}{1} 
\setcounter{equation}{0}

\subsection{Self-duality of the Aubry-Andr\'{e} model}
In this subsection, following the results from Ref.~\cite{aubry1980}, we describe the essence of the self-duality argument and how it can be used to analyze the localization properties of the eigenmodes. The model we consider first is the tight-binding version of the Aubry-Andr\'{e} model with the Hamiltonian
\begin{align} \label{AA_ham}
t \, ( \psi_{j+1} + \psi_{j-1} ) + \lambda \,  \cos(2\pi b j + \phi) \, \psi_{j} = E \, \psi_j \, ,
\end{align}
where $\psi_j$ is the wavefunction at site $j$, $t$ the nearest-neighbour hopping amplitude, and $\lambda$ the amplitude of the on-site potential modulation. We transform the wave functions $\psi_j$ as
\begin{align}	\label{fourier_transform}
\psi_j = e^{i \theta j} \sum_{k=-\infty}^{k=\infty} f_k e^{i k (2\pi b j + \phi) }
\end{align}
and obtain the Fourier-transformed equation 
\begin{align}  \label{dual_eq}
\frac{\lambda}{2} \, ( f_{k+1} + f_{k-1} ) + 2t \,  \cos(2\pi b k + \theta) \, f_{k} = E \, f_k \, .
\end{align}
The two equations [\eqref{AA_ham} and \eqref{dual_eq}] are identical at the critical point, i.e. if $\lambda/ t = 2$. Now, we note that if we find a localized solution in Fourier space, $f_k$, such that $\sum_k |f_k|^2 < \infty$ then, if \eqref{fourier_transform} converges, the solution of Eq.~\eqref{AA_ham} has the property that $\sum_n |\psi_j|^2 = \infty$. In other words, the transformation \eqref{fourier_transform} exchanges the localization properties of $\psi$ and $f$ eigenmodes, namely if $\psi$ is extended, $f$ is localized and vice-versa. 
In the limit when $\lambda / t \rightarrow 0$ (e.g, when $\lambda \rightarrow 0$), Eq.~\eqref{AA_ham} describes a metallic chain with all modes $\psi$ being extended, while when $\lambda / t \rightarrow \infty$ (e.g, when $t \rightarrow 0$), the hopping term is negligible and the eigenmodes are localized on one site. Since the same argument is applicable also for Eq.~\eqref{fourier_transform}, we conclude that the transition happens exactly at critical point $\lambda/ t = 2$.

\newpage
\subsection{Generalized self-duality of interpolating Aubry-Andr\'{e}-Fibonacci (IAAF) model}
We apply the self-duality argument presented in the previous subsection to the IAAF model [Eq.~(1) in the main text], while taking a small $\beta$ expansion. We obtain the critical line $\lambda_{C}/t$ that bounds the phase where all eigenmodes of the model are extended.
First, we expand the potential modulation [cf. Eq.~(2) in the main text]  for small $\beta$-s and obtain
\begin{align}
V(x, \beta) &= \chi + \frac{1}{3} \chi \, (1-\chi^2) \, \beta^2 + \mathcal{O}(\beta^3),
\end{align}
where $\chi \equiv \cos(2 \pi b x + \phi) -\cos(b \pi)$. Notice that we use the continuous version of the potential defined in the main text. To return to the discrete version, we restrict the position $x$ to be a set of integer numbers.
After expanding the potential, we approximate the quadratic $\beta$-term as
\begin{align}
V_j(\beta) &\approx \chi + \frac{1}{3} U \, \beta^2 \, \chi,
\end{align}
where $U$ is the spatial average over a single period of the potential modulation $V(x, \beta)$ 
\begin{align}
U &= b \int_{0}^{b^{-1}} {\rm d} x \, [1-\chi^2] =\\
&= b \int_{0}^{b^{-1}} {\rm d} x \, [1-(\cos(2 \pi b x + \phi) -\cos(b \pi))^2]=\\
&= -\frac{1}{2} \cos(2\pi b) \, .
\end{align}

\begin{figure}[t!]
	\center
	\includegraphics[scale=1]{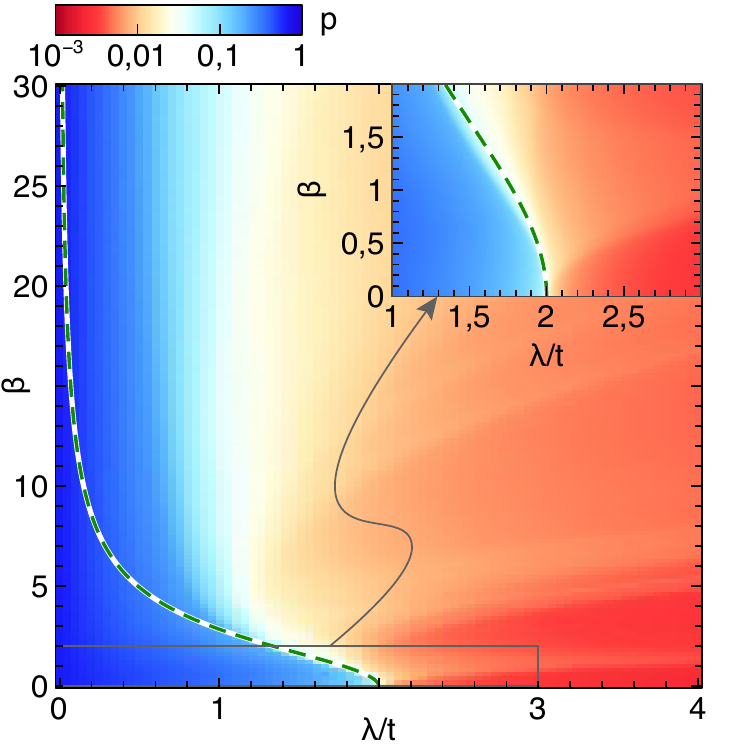}
	\caption{
		\textbf{ The localization phase diagram averaged over the whole spectrum. }
		The average localization phase diagram for all states in the spectrum obtained using the mean participation ratio [Eq.~\eqref{PR}]. The green dashed line marks the analytical result from the generalized self-duality argument~\eqref{lambda_critical}. The line separates extended (blue) from localized (red) phase. The structure of localization lobes present in Fig.~1 from the main text and in Supplementary Section~2 is washed out due to the averaging over all eigenmodes of the spectrum.
		The size of the system is $L=610$ sites.
		\label{fig1}}
\end{figure}	
In this approximation, the effective potential remains to be a cosine function incommensurate with the underlying lattice, but its amplitude is altered with the tuning parameter $\beta$. Therefore, the Hamiltonian (Eq.~(1) from the main text) keeps the same shape
\begin{equation} 
\mathcal{H} \, \psi_j = t \, ( \psi_{j+1} + \psi_{j-1} ) + \Lambda \,  \left[ \cos(2 \pi b j + \phi) -\cos(b \pi) \right] \, \psi_{j}\,,
\label{ham}
\end{equation}
but with $\Lambda$ which is now a function of $\beta$
\begin{align}
\Lambda= \lambda \, \left( 1-\frac{1}{6}\cos(2\pi b)\,\beta^2 \right). 
\end{align}
The self-duality point of the model effective [Eq.~\eqref{ham}] is $\Lambda / t= 2$, which implies the critical line
\begin{align}
\frac{ \lambda_{C} }{t} = \frac{2}{\left( 1-\frac{1}{6}\cos(2\pi b)\,\beta^2 \right) } \, . 
\label{lambda_critical}
\end{align}
The line $\lambda_{C}/t$ separates the extended phase from the localized one, as shown in Supplementary Fig.~\ref{fig1}:
The validity of our approximation, together with its meaning can be seen if we plot the phase diagram using the mean participation ratio defined via the inverse participation ratio (IPR) from the Eq.~(3) in the main text as~\cite{Thiem_2013}
\begin{align}
p = \left\langle IPR^{-1} \right\rangle.
\label{PR}
\end{align}
Brackets $\left\langle ... \right\rangle$ denote the arithmetic average over all eigenmodes in the spectrum. If all the eigenmodes are localized on one site, $p \rightarrow 0$, and otherwise, for extended states, $p \rightarrow 1$.

\newpage
\section{II. Phase diagram for higher-energy states}
\setcounter{enumi}{1} 
\setcounter{equation}{0}
In Supplementary Figs.~\ref{fig2}a-f, we plot the localization phase diagram for different states in the spectrum. The same structure of lobes of localized modes, separated by narrow slivers where the mode is extended states, can be observed. Furthermore, as it is shown in Supplementary Fig.~\ref{fig2}g. all states have similar stepwise cascade of IPR when $\beta$ is tuned for constant $\lambda/t \gg 2$. 
This indicates that all states reach criticality for $\beta \rightarrow \infty$ with the same mechanism as discussed in the main text.
\begin{figure}[h!]
	\center
	\includegraphics[scale=1]{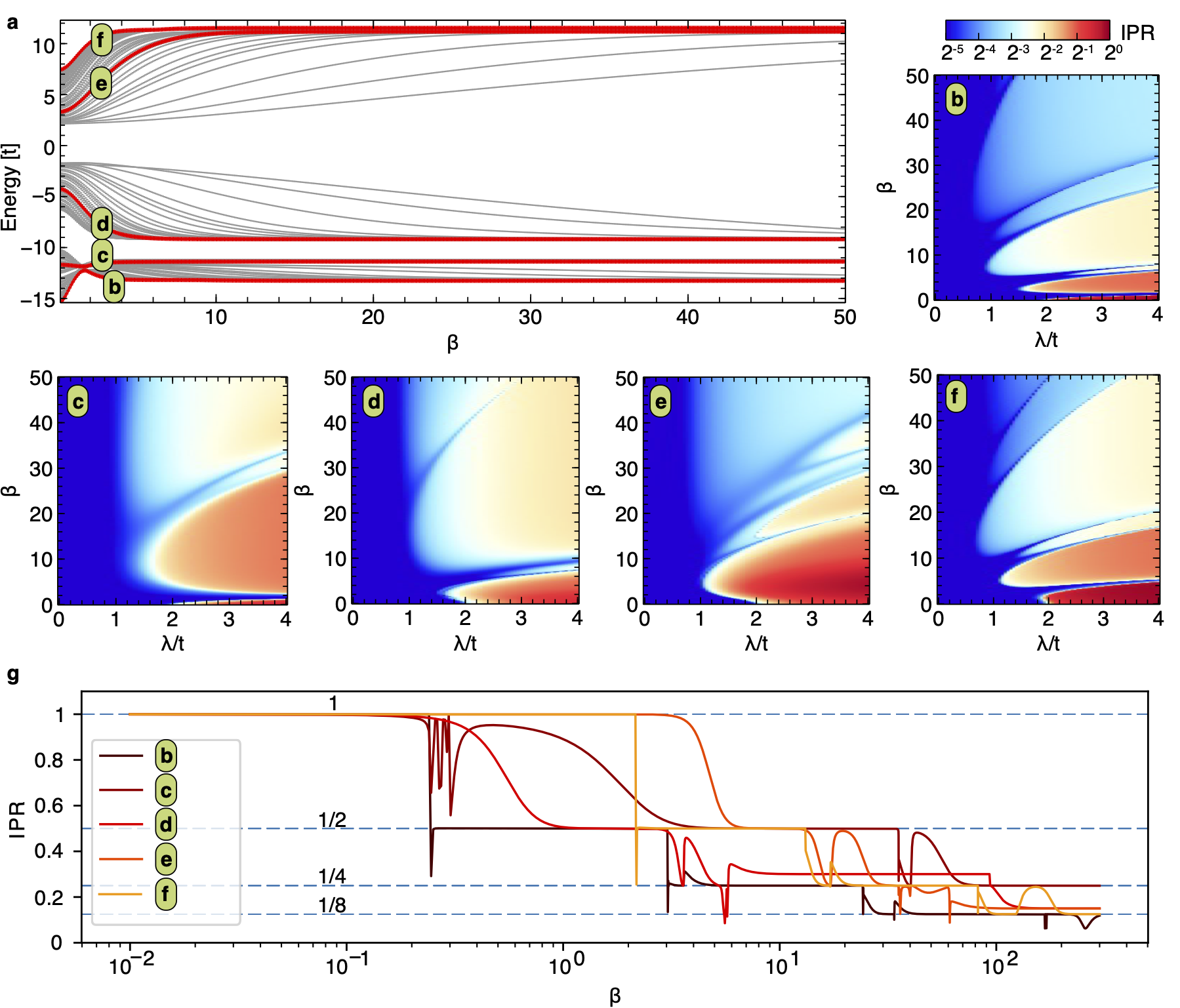}
	\caption{
		\textbf{ The localization phase diagram for several states in the spectrum. }
		\textbf{a}, 
		The energy spectrum as a function of $\beta$ for $\lambda/t=5.5$. Several states are emphasized (red),
		for which we show in \textbf{b}-\textbf{f}, the IPR localization phase diagram [cf.~Eq.(2) and Fig.~1c in the main text). 
		\textbf{g},
		IPR as a function of $\beta$ for the states marked with red in \textbf{a}. All of them show similar stepwise cascade with height decreasing with a factor of 2. Here, we use $\lambda / t = 100$. A system of length $L=144$ sites is used for all plots. 
		\label{fig2}}
\end{figure}

\newpage
\section{III. Two-site localization mechanism}
\setcounter{enumi}{1} 
\setcounter{equation}{0}
In Supplementary Fig.~\ref{fig3}, the mechanism behind the relocalization on 2 sites in explained. In Supplementary Fig.~\ref{fig3}a, we see that at some point along the chain, the AA potential modulation [see Eq.~(2) from the main text] arranges the onsite energies such that a nearest-neighbor pair appears close in energy. Such states strongly hybridize due to the finite hopping strength $t$ (see Supplementary Fig.~\ref{fig3}b). With increasing $\beta$, the marked pair of states moves towards lower energies, goes through the delocalization transition, and overtakes the role of the lowestmost energy eigenmode (see Supplementary Fig.~\ref{fig3}c and \cite{supmat_video_1}). The lowest energy eigenmode is then localized on two sites.
\begin{figure}[h!]
	\center
	\includegraphics[scale=0.9]{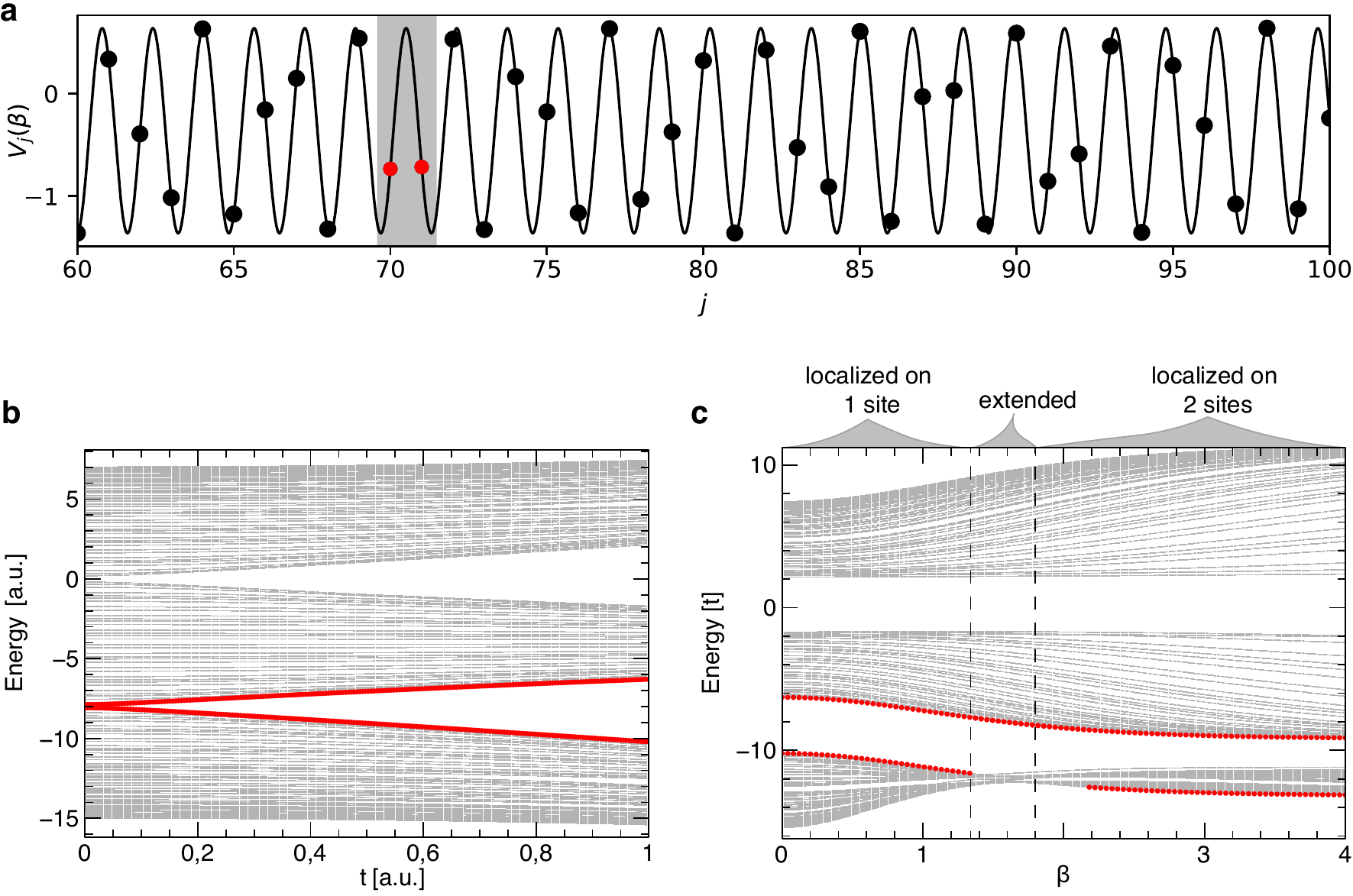}
	\caption{
		\textbf{ Mechanism of localization on two sites. }
		\textbf{a}, 
		The potential modulation [cf.~Eq.~(2) in the main text] for $\beta=0$, i.e., the AA modulation. We mark the pair of states that are both closest in energy and nearest neighbors in space, which will eventually become the lowest energy mode once $\beta$ is increased.   
		\textbf{b},
		Energy spectrum of the AA model as a function of $t$ for $\beta=0$, $\lambda=5.5$. With red lines, we track the aforementioned pair of states marked in \textbf{a}. For $t > 0$, a gap opens and the two marked states lie at the band edges.
		\textbf{c},
		Energy spectrum of the IAAF model as a function of $\beta$ with the marked pair of states as in \textbf{a,b}. We take $t=1$ and the marked states are hybridized already for $\beta=0$. By increasing $\beta$, these hybridized states are moving towards lower energies, and at some point overtake the role of the lowest energy state in the system (see also Supplementary Video~1). Vertical dashed lines mark three regions, where the lowest energy eigenmode is: localized on 1 site, extended or localized on two sites. 
		In all plots, system has $L=144$ sites.
		\label{fig3}}
\end{figure}

\subsection*{Supplementary Video~1 and 2}

Supplementary Video~1.	\textbf{ The mechanism of localization on two sites }
(Left panel) The grey horizontal lines mark the density of states as a function of energy and site number $j$, calculated for the IAAF model [cf. Eq.~(1) from the main text] at different values of $\beta$. Blue line marks the potential modulation [cf. Eq.~(2) in the main text] with discrete energies (blue circles). We mark the lowest-energy eigenmode at $\beta=0$ (blue square) and hybridized pair of nearest neighbours (red square). 
(Right panel) Spectrum of the IAAF model as a function of $\beta$. We mark the same states as in the left panel and with the same color code.
Here, we use $\lambda/t=10$ and L$=34$ sites.

\bigskip
Supplementary Video~2.	\textbf{ The mechanism of localization on four sites }
(Upper left panel) The grey horizontal lines mark the density of states as a function of energy and site number $j$, calculated for the IAAF model [cf. Eq.~(1) from the main text] at different values of $\beta$. Blue line marks the potential modulation [cf. Eq.~(2) in the main text] with discrete energies (blue circles). We mark the hybridized pair of nearest neighbours (red square) and states which hybridize on four sites (green squares) at high $\beta$-s. 
(Upper right panel) Spectrum of the IAAF model as a function of $\beta$. We mark the same states as in the left panel and with the same color code.
(Lower panel) Enlarged region from the upper right panel to see how the mode hybridized on four sites (green square) overtakes the role of the lowest-energy mode.
Here, we use $\lambda/t=10$ and L$=34$ sites.

\newpage
\section{IV. Additionnal information on the continuum model}
\setcounter{enumi}{4} 
\setcounter{equation}{0}

The eigenmodes in the continuum model are obtained numerically by diagonalization of the nearly-free particle Hamiltonian, given by Eq. (4) in the main text.
The parameter values are extracted from the experiment: we use a polariton mass $m = 3 \times 10^{-5} m_e$, with $m_e$ the free electron mass, and step length $a = \SI{2}{\micro m}$.

The definition of the IPR, given in the tight-binding model by Eq. (3) in the main text, needs to be adapted to the continuum model. In the latter case, we define the IPR of a mode $\psi(x)$ as:
\begin{align}
\mathrm{IPR} = a \int |\psi(x)|^4 \mathrm{d}x
\label{eq:IPR}
\end{align}
The values of the IPR presented in Fig. 2a-c of the main text are computed using this definition.

\section{V. Kinetic energy and AA localization in the continuum model}
\setcounter{enumi}{5} 
\setcounter{equation}{0}

Let us comment here on the relevant kinetic energy scale in the lowest band in the continuum model. In a periodic system, the characteristic kinetic energy scale is given by the recoil energy $E_R = \hbar^2 k_R^2/ 2 m$, where $k_R = \pi / a$ is the edge of the first Brillouin zone (for a unit cell of size $a$). In a quasi-periodic chain, no Brillouin zone can be defined. However, in the case of the IAAF model, the gap labeling theorem predicts that gaps in the energy spectrum open at very specific wavevectors $k_{p, q}$ uniquely identified by two integers $(p,q)$~\cite{Belissard1992}:
\begin{equation}
k_{p, q} = \frac{\pi}{a} \left(p + bq \right)
\label{eq:gap_labeling}
\end{equation}
We remind that $b$ is the inverse of the golden mean. For our chosen value of letter size $a = \SI{2}{\micro m}$, the main gap above the lowest band opens at $k = \SI{0.6}{\micro m^{-1}}$, corresponding to $(p, q) = (-1, 1)$. In analogy with the recoil energy in a periodic system, we get a characteristic kinetic energy scale $E_R = \SI{0.47}{meV}$. Thus, in Fig. 2a of the main text, the localization transition in the AA limit is observed for $\lambda_{\rm eff} \approx \SI{1}{meV} = 2.2 E_R$. This is consistent with the value $\lambda / t = 2$ for the AA localization transition in the tight-binding model, for which the relevant kinetic energy scale is $t$.

\section{VI. Transverse profile of modes in different 1D subbands}

\begin{figure}[h]
	\includegraphics[width=0.7\linewidth]{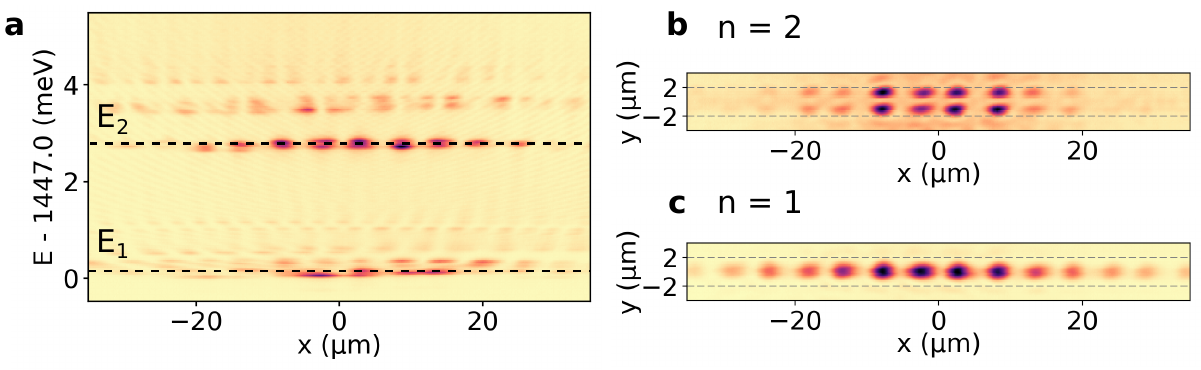}
	\caption{\label{figS_n12} \textbf{2D imaging of the eigenmodes in different subbands.} 
		\textbf{a} PL intensity measured as a function of position $x$ and energy for a wire with $\beta = 0$ and $\lambda_1 = \SI{0.2}{meV}$, reproduced from Fig. 2.g of the main text.
		\textbf{b, c} 2D image of the eigenmodes in the same wire, obtained by spectrally filtering the PL emission at energy \textbf{b}: $E_2$ and \textbf{c}: $E_1$. Dashed gray lines indicate schematically the edges of the wire (for simplicity, lateral modulations are not shown).
	}
\end{figure}

Real-space images of eigenmodes in the $n = 1$ and $n = 2$ subbands can be obtained by spectrally filtering the PL emission at the energy of each band. The 2D map of the emission pattern is then reconstructed from spectra such as the one shown in Supplementary Fig.~\ref{figS_n12}a, measured at different values of the lateral position $y$ on the wire. The results for both $n=1$ and $n=2$ subbands are presented in Supplementary Fig.~\ref{figS_n12}b,c. We recognize the nature of the $n = 1$ and $n = 2$ modes discussed in the main text: the $n=1$ modes lateral profile has a single bright lobe, while $n = 2$ modes have the characteristic transverse profile with two bright lobes, and a zero at the center of the wire.

\section{VII. Estimation of the band curvature for $\beta = 1$, $\lambda_2 = \SI{2.4}{\meV}$}

\begin{figure}[h]
	\includegraphics[width=0.8\linewidth]{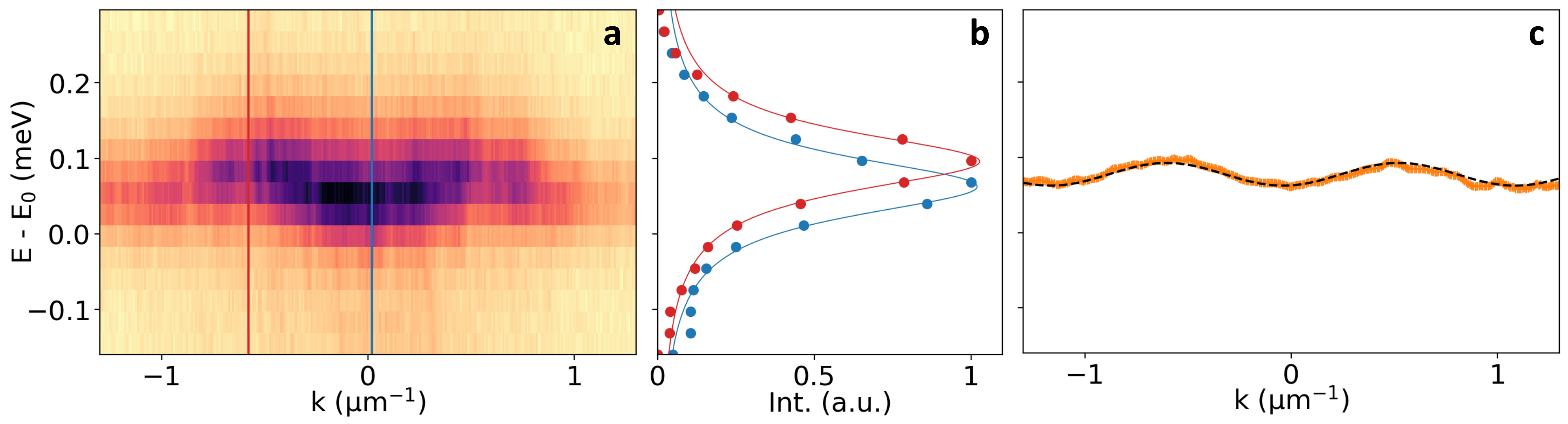}
	\caption{\label{figS_beta1} \textbf{Estimation of band curvature for $\beta = 1$, $\lambda_2 = \SI{2.4}{\meV}$.} 
		\textbf{a,} Zoom on the lowest $n=2$ band in momentum space, reproduced from Fig.~3b of the main text.
		\textbf{b,} (Dots) Cut in the spectrum at two different values of $k$, corresponding to the solid lines in \textbf{a}, and (solid lines) respective Lorentzian fits.
		\textbf{c,} Energy of the band versus $k$, fitted by a cosine function (dashed black line) with amplitude $\SI{30}{\micro \eV}$.
	}
\end{figure}

The existence of delocalized modes in the case $\beta = 1$, $\lambda_2 = \SI{2.4}{\meV}$ is evidenced by the presence of a band with finite curvature, as seen in the inset of Fig. 3.b$_1$ in the main text. In Supplementary Fig.~\ref{figS_beta1}, we show that in this case, the band is well fitted by a cosine function, attesting the existence of long-ranged coherence. For each $k$ value, the energy spectrum is fitted with a Lorentzian profile with central energy $E_b(k)$, as represented in Supplementary Fig.~\ref{figS_beta1}.b. The extracted values of $E_b(k)$ are reported in Supplementary Fig.~\ref{figS_beta1}.c, together with the cosine fit. Note that the value of the band width extracted from the fit is approximately $\SI{30}{\micro \eV}$, i.e., below the polariton linewidth and comparable with the resolution of the spectrometer. This explains why the value of $\Delta k$ remains bigger than the one measured for small $\lambda$ values in Fig.~3d of the main text.

\section{VIII. Numerical simulations for Figs. 3a-d from the main text}

In Supplementary Fig.~\ref{figS_sim1D} presents numerical calculations of the local density of modes, in the continuum model, that correspond to the measured spectra presented in Figs. 3a-d of main text. 
An excellent agreement is found with the experiment: note, for example, how the modes localized on a single letter (down arrow) and on two letters (up arrow), located at the same position as in the experiment, and exchange energy across the delocalization-transition. The delocalization occurs when the energies of the one- and two-letter modes become resonant, i.e., between $\beta = 0.8$ and $\beta = 1$.

Note that in the derivation of the 1D piecewise potential $U(w_j)$ approximations were taken that neglect corrections from the lateral confinement~\cite{Tanese2014}. Hence our numerical simulations are computed using a value of the modulation amplitude $\lambda_{\rm eff} = \SI{1.8}{meV}$, that is slightly smaller than the nominal value $\lambda_{\rm eff} = \SI{2.4}{meV}$ reported in Fig. 3 of the main text. For this nominal value, the delocalization transition is observed around $\beta \approx 1.1$, as seen in Fig.~2c of main text.

\begin{figure}[h]
	\includegraphics[width=0.8\linewidth]{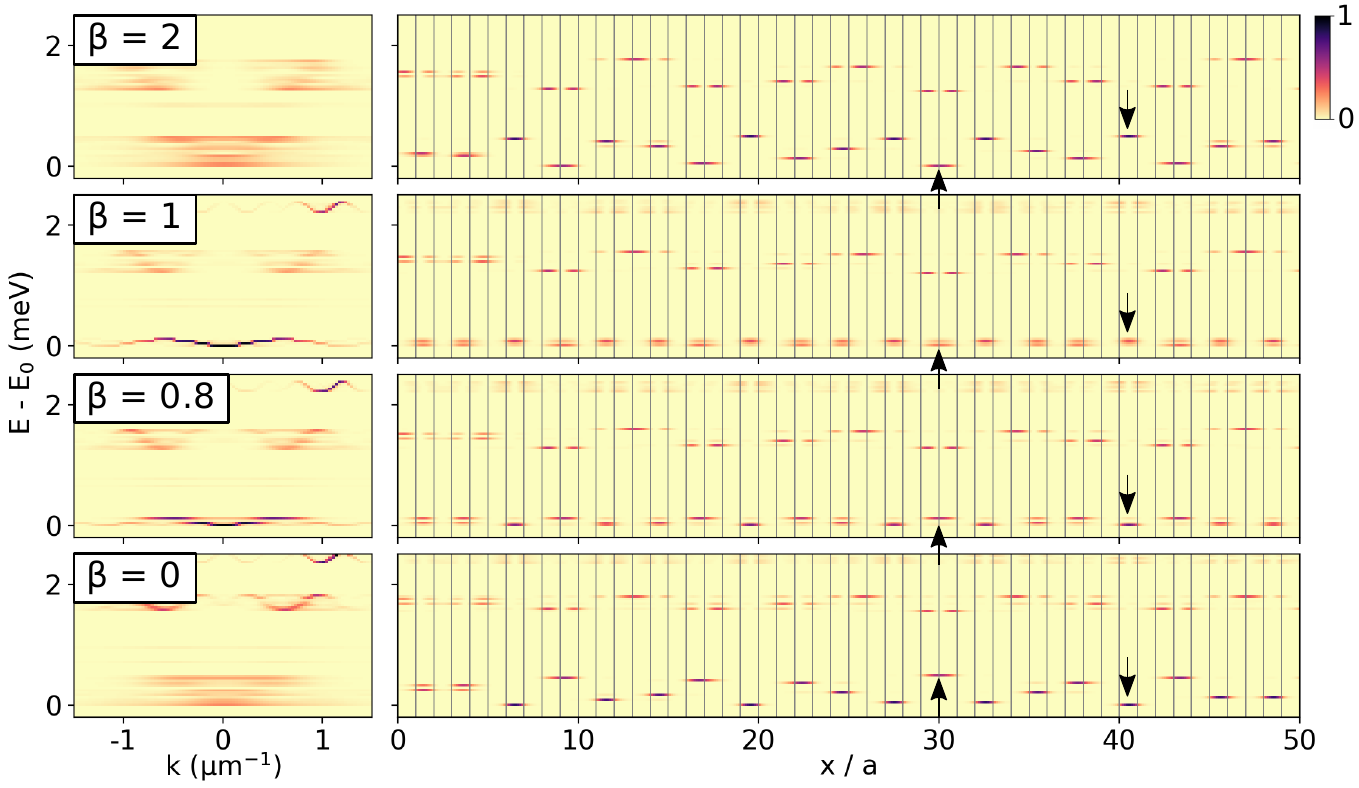}
	\caption{\label{figS_sim1D} \textbf{Numerical simulations corresponding to Fig. 3.a-d of main text.} 
		\textbf{a-d} Calculated local density of modes, in (left) momentum- and (right) real-space, for wires with $\lambda_{\rm eff} = \SI{1.8}{\meV}$ and \textbf{a:} $\beta = 2$ ; \textbf{b:} $\beta = 1$ ; \textbf{c:} $\beta = 0.8$ and \textbf{d:} $\beta = 0$. Colorbar marks the normalized intensity.
	}
\end{figure}

\end{document}